\documentclass[letterpaper]{article}
\usepackage{natbib,alifeconf,float, textcomp, hyperref}
\hypersetup{
    colorlinks=true,
    linkcolor=blue,
    filecolor=blue,      
    urlcolor=blue,
    citecolor=blue,
}

\title{Coloured noise time series as appropriate models for environmental variation in artificial evolutionary systems}
\author{Matt Grove$^{1}$, James M. Borg$^{2}$ \and Fiona Polack$^{2}$
\mbox{}\\
$^1$Department of Archaeology, Classics and Egyptology, University of Liverpool, UK \\
$^2$School of Computing and Mathematics, Keele University, UK \\
Matt.Grove@liverpool.ac.uk, j.borg@keele.ac.uk, f.a.c.polack@keele.ac.uk}
\begin{document}

\maketitle
\begin{abstract}
Ecological, environmental and geophysical time series consistently exhibit the characteristics of coloured ($1/f^\beta$) noise. Here we briefly survey the literature on coloured noise, population persistence and related evolutionary dynamics, before introducing coloured noise as an appropriate model for environmental variation in artificial evolutionary systems. To illustrate and explore the effects of different noise colours, a simple evolutionary model that examines the trade-off between specialism and generalism in fluctuating environments is applied. The results of the model clearly demonstrate a need for greater generalism as environmental variability becomes `whiter', whilst specialisation is favoured as environmental variability becomes `redder'. Pink noise, sitting midway between white and red noise, is shown to be the point at which the pressures for generalism and specialism balance, providing some insight in to why `pinker' noise is increasingly being seen as an appropriate model of typical environmental variability. We go on to discuss how the results presented here feed in to a wider discussion on evolutionary responses to fluctuating environments. Ultimately we argue that Artificial Life as a field should embrace the use of coloured noise to produce models of environmental variability.
\end{abstract}

\section{Introduction}
Empirical measurement has shown that time series ranging from environmental temperatures and precipitation levels to population sizes, earthquake frequencies, and historical river levels are all well characterised by coloured ($1/f^\beta$) noise \citep{Halley1996, Inchausti2002}, and that simple Gaussian `white noise' provides an insufficient null hypothesis for the noise component of a signal in the majority of cases \citep{Groth&Ghil2015}.

\begin{figure}[h]
    \centering
    \includegraphics[width=\linewidth]{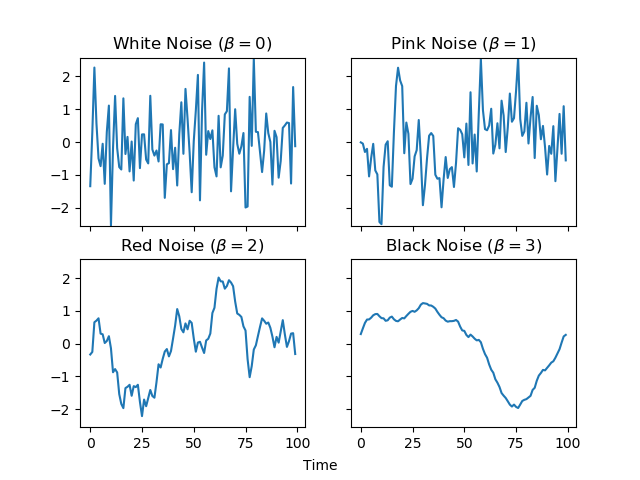}
    \caption{Examples of time series exhibiting white noise (top left), pink noise (top right), red/Brownian noise (bottom left), and black noise (bottom right). As $\beta$ increases the level of temporal autocorrelation also increases. \cite{Halley1996} and \cite{Halley2004b} suggests that pink noise (aka. $1/f$ noise) is an appropriate model of typical environmental and ecological noise}
    \label{fig:NoisePlots}
\end{figure}

When time series are decomposed into constituent sine waves via Fourier transforms and plotted as $log(frequency)$ against $log(power)$,  coloured noise signals are well approximated by straight lines of the form ($1/f^\beta$), with ($f$) equal to frequency and ($\beta$) used to describe the colour of the noise. As $\beta$ grows, the noise is said to `redden', from white noise ($\beta=0$), through pink nose ($\beta=1$), to red/Brownian noise ($\beta=2$). Values of $\beta$ above 2 result in black noise, with values of $\beta$ below $0$ characterised as blue noise. Figure \ref{fig:NoisePlots} provides examples of white to black noise time series; from these examples we can see that as $\beta$ increases the level of autocorrelation (correlation of adjacent values of the time series) also increases. 

Artificial Life models used to study evolutionary dynamics in areas such as the evolution of intelligence, morphology, and social behaviour frequently include noise parameters, or some form environmental variability. For example, environmental variability has been used in evolutionary robotics to explore the evolution of robust controllers \citep{Jakobi1995, Stanton2018}; in exploration of the emergence of social, cultural and plastic adaptation \citep{Channon1998,Jones2011, Borg2011, Borg2012, Grove2014, Jolley2016, Grove2018, Bullinaria2018}; in evolving digital organisms to explore the evolution of evolvability \citep{Canino-Koning2016}; and in exploration of complex and noisy environments for the emergence of open ended evolution \citep{Channon1998, Channon2019}. The time series applied in such Artificial Life research tend to be either cyclical, fluctuating or pulsing. Where noise is deliberately added to a system, it tends to be Gaussian `white noise'. As such, the nature of the environmental variability or noise employed in Artificial Life research rarely accords with what we know of the empirical characteristics of real environmental, ecological or geophysical noise. 

Here, we take inspiration from ecological, geophysical, and evolutionary time series, which have been shown to exhibit coloured noise at the red end of the spectrum ($0\leq\beta\leq2$). Examples include temperature (both marine and terrestrial), precipitation, river levels and meanders \citep{Mandelbrot1969, Steele1985, Cuddington1999, Vasseur2004}, and ecological phenomena such as population growth, population persistence, and extinction \citep{Miramontes1998, Inchausti2001, Inchausti2002, Halley2004b}. It has been suggested that marine environmental noise (such as sea temperature) tends to be `redder' than terrestrial environmental noise \citep{Vasseur2004}, whilst some geophysical phenomena exhibit `black' noise with $\beta>2$ \citep{Mandelbrot1969, Cuddington1999}. Animal population evolutionary dynamics typically exhibit $1/f$ `pink' noise at a level that appears to be greater than would be expected from the environmental noise that the population was exposed to \citep{Inchausti2002}.

\cite{Halley1996} identifies $1/f$ `pink' noise as an appropriate model of typical environmental noise. He notes that ecologists expect evolution to be affected both by rare and common events; white noise, $1/f^0$, contains all frequencies but lacks autocorrelation, and is not an appropriate model for a real environment, because it under-represents rare but significant disruptive events. Similarly, reddened noise ($1/f^{\beta}$, $\beta$ tending to 2) over-emphasises longer-term correlations. By contrast, in pink noise power decays as approximately the inverse of frequency, giving appropriate weight to both common and rare environmental events. It therefore follows not only that Artificial Life researchers should be interested in reddened noise, as opposed to white or uncoloured noise, but also that they should direct their attention to pink noise in order to produce ecologically inspired models of environmental variability.

\section{Coloured Noise, Population Persistence and Evolutionary Dynamics}

A particular focus of coloured noise research in recent years has been evolutionary changes in population size (growth, extinction). Using the Ricker model \citep{Ricker1954} which provides a more realistic version of the conventional logistic model of population growth, \cite{Ripa1996} show that redder noise lessens the risk of extinction, concluding that the autocorrelation characteristics are critical in determining whether a population grew (red noise) or declined, ultimately to extinction (blue noise). These simulations demonstrate that noises of different colours have profoundly different effects on population dynamics, and that we must look beyond the traditional Gaussian or white noise models to establish realistic models of environmental variability.

Similar findings were obtained subsequently by \cite{Cuddington1999}, using a variant of the Ricker model that explicitly includes coloured noise \citep{Petchey1997}. This research extended noise analysis to colours $1/f^{\beta}$, $\beta\leq3.2$. High $\beta$ values are shown to be associated with increasingly long population persistence times compared to less reddened noise ($0.5\leq\beta\leq1$). The persistence of the population is shown to be very robust with respect to the form of the underlying model. Again, \cite{Cuddington1999} note that white noise, with its lack of autocorrelation, does not capture the features of the natural environments within which evolution occurs.

\cite{Cuddington1999} present a method for modelling of single-species population dynamics, which provides a robust basis for analysis of noise effects. Using the variant Ricker model \citep{Petchey1997}, these authors add a stochastic element derived from the population size at a given time ($N_t$) by drawing from a Poisson distribution, $Z$, with the mean given by the expected model output. $1/f^\beta$ noise with spectral exponents ranging from $0$ to $3.2$ is generated using a spectral synthesis approximation which applies an inverse fast Fourier transform to amplitudes and periods with the desired spectral exponent.  

An intuition based on these findings is that evolution in an environment that is dominated by Brownian-style noise produces robust populations that can adapt effectively to potentially significant but rare environmental challenges and events. This intuition is shared by \cite{Halley1999}, who note that the effect of reddening is to increase the variance observed in longer time series, but that, contrary to the expectation that populations are more likely to become extinct in more extreme environments, evolution in such conditions is more robust to environmental variance. 

Inchausti and Halley (\citeyear{Inchausti2001},\citeyear{Inchausti2002},\citeyear{Inchausti2003}) again confirm increasing variability over time using time series of 30 years or more (max 157 years) from the Global Population Dynamics Database, for 544 animal populations of 123 species. They again note the apparent contradiction between red-noise environments and population robustness. They note that a non-regulated population may benefit by being able to wander to an arbitrarily high, and thus relatively invulnerable, population level, whereas the abundance of a strongly-regulated population is constrained by density-dependent interactions, leaving a population that is vulnerable to sudden extinction-triggering events. The contradictory relationship between red-noised environments and populations is evaluated by \cite{Morales1999}, who points out that whilst extinction risk can be increased in redder noised environments, these results are only seen when noise is applied to population growth, whereas when noise instead dictates some ecological phenomena (such as the carrying capacity of the environment), redder noise actually leads to decrease extinction rates. From these and wider ecological time series, \cite{Halley2004b} subsequently confirm the important characteristics of pink noise: its proportional power across the frequency range, its long memory, and its non-stationarity. 

\subsection{Generalism, specialism and sensitivity in fluctuating environments}
Recent work by \cite{vanderBolt2018} has begun to shed some light on the dangers of climate reddening, where elevated levels of temporal autocorrelation (where noise transitions beyond pink noise) can lead to the persistence of anomalous climatic events or trends. This is demonstrated by both the red and black noise time series in Figure \ref{fig:NoisePlots}, where the environmental state can persist well outside of the mean environmental state in a way not seen under white or even pink noise. Under elevated levels of temporal autocorrelation, \cite{vanderBolt2018} demonstrates that the chances of critical transitions in climate-sensitive systems increases, leading to the potential loss of climate-sensitive systems such as coral reefs, ice sheets, and forests. \cite{vanderBolt2018} (echoing the work of \cite{Halley2004b}) goes on to conclude that understanding `climate memory' is as important as understanding variability when researching climate change, and the implications of climate change on climate-sensitive systems.

We believe this sentiment should also be carried forward in to evolutionary dynamics research; evolution reacts to environmental change over multiple time scales, hedging its bets against both short term and long term environmental change in order to not only maximise fitness in the short term, but also increase the long term likelihood of survival. Recent work by \cite{Haaland2019} and \cite{Haaland2020} demonstrates the need for evolution to select for individuals that are  ``more generalist than required to simply maximize their own expected fitness'' when environments fluctuate. Whilst neither \cite{Haaland2019} nor \cite{Haaland2020} applies coloured noise when producing variable environments, both show a strong dependency between the level of plasticity exhibited by individuals and the unpredictability of the environment. As real environments exhibit coloured noise, and therefore autocorrelation and `memory', it seems wise for us to explore and better understand evolutionary dynamics under conditions determined by coloured noise. 

\section{Grove's model of adaptive behaviour}

To explore the effects of different noise colours further, we adapt a model originally developed by \cite{Grove2014} to analyse the trade-off between specialism and generalism in fluctuating environments. The original research employed sine-wave environments of varying frequency and amplitude, but here we instead use a range of coloured noises in the white ($1/f^0$) to red ($1/f^2$) spectrum to examine their effects on the levels of tolerance (\texttildelow generalism) that evolve in simulated populations.

The model employed is as per \cite{Grove2014}, with the only modifications being that mutation is now governed by a Gaussian operator ($N(0,0.1)$) and reproduction is altered so as to be asexual; the latter change makes minimal difference, but is computationally faster. The substantial advance of the current research is that it employs coloured noise environments rather than the simple sine waves used in the earlier paper. A population of 1,000 agents with two loci corresponding to the mean and standard deviation of a normal distribution have their fitness assessed each iteration according to a Gaussian function (see equation \ref{eq:Gaussian}). 

\begin{equation} \label{eq:Gaussian}
   f(a_{i,t}) = \frac{1}{\sigma_{i,t}\sqrt{2\pi}}e^{-\frac{1}{2}(\frac{E_t-\mu_{i,t}}{\sigma_{i,t}})^{2}}
\end{equation}

Where $f(a_{i,t})$ denotes the fitness of agent $i$ at time $t$, $E_t$ denotes the environmental value at time $t$, and $\mu_{i,t}$ and $\sigma_{i,t}$ denote respectively the values at the `mean' and `standard deviation' loci of agent $i$ at time $t$. Determining fitness in this way is broadly consistent with other recent work concerned with the analysis of the trade-off between specialism and generalism in fluctuating environments \citep{Haaland2019,Haaland2020}.

Each iteration the least fit 500 individuals are removed, to be replaced with 500 individuals sampled from the fittest 500 individuals via fitness-proportionate selection. The 500 new individuals are then mutated slightly at each locus according to the mutation operator ($N(0,0.1)$). Mean $\mu$, $\sigma$, and fitness values for the population are recorded before proceeding to the next iteration. We reflect below on how changes to the variance of the mutation operator and the proportion of agents replaced each iteration affect the results presented.
\begin{figure*}[h!]
    \includegraphics[width=\textwidth]{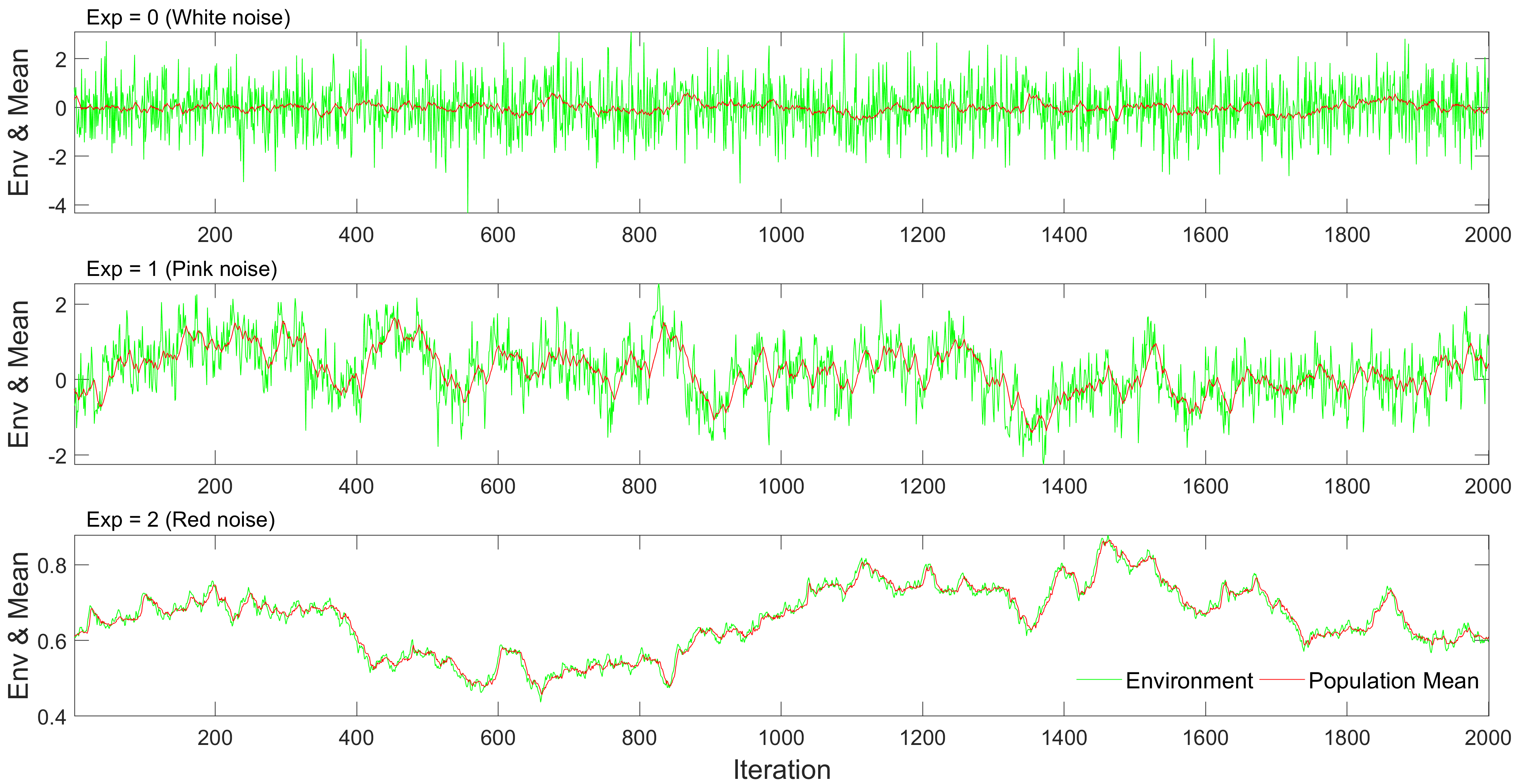}
    \caption{Three 2,000-iteration snapshots of single model runs. Agents are better able to track `reddened' noise, whereas the trajectory for white noise suggests that agents evolve towards the running mean of the environmental series and couple this with a broader environmental tolerance.}
    \label{fig:Snapshot}
\end{figure*}

\begin{figure}[h!]
    \includegraphics[width=\linewidth]{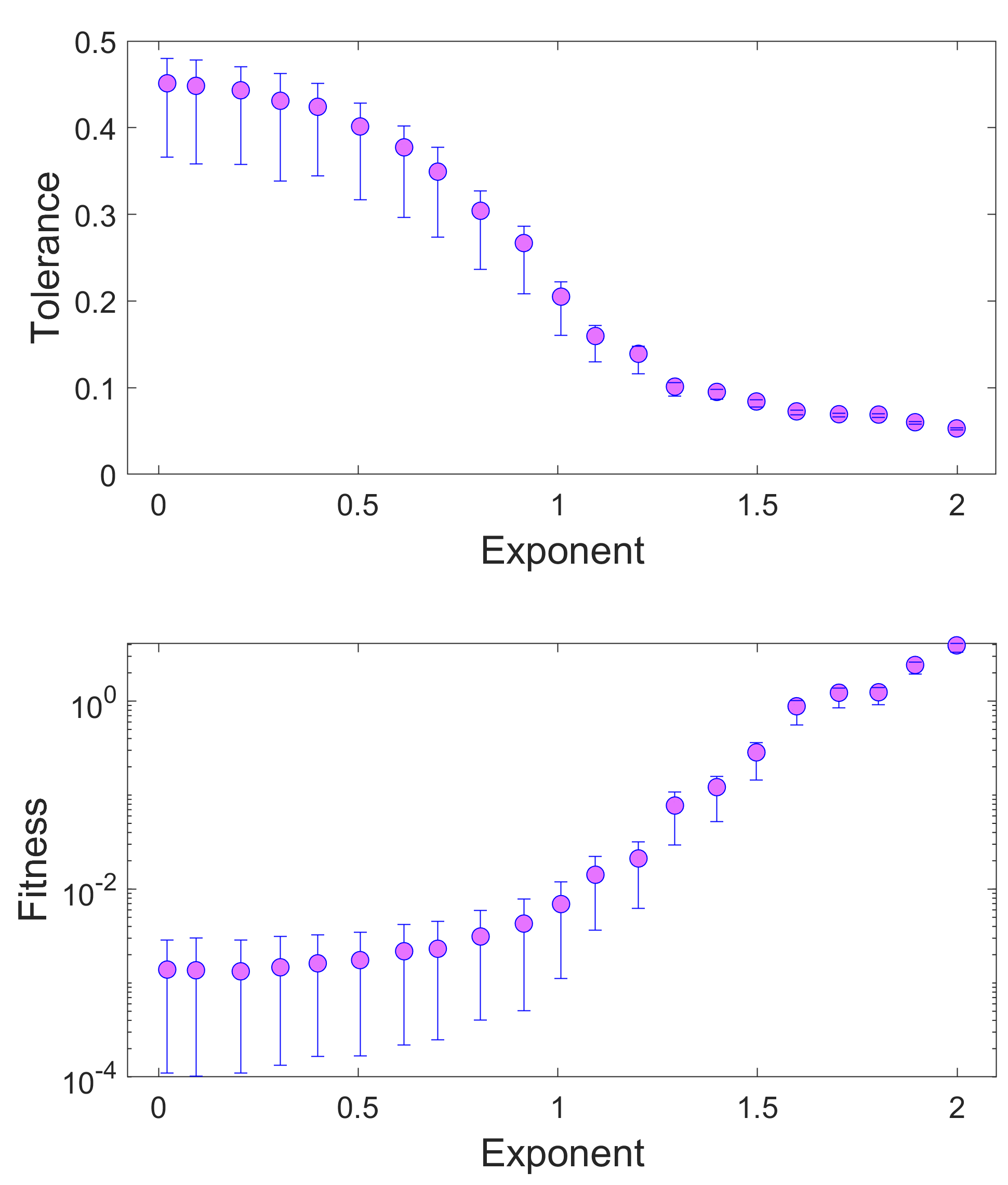}
    \caption{Tolerance and fitness values for populations evolving in environments characterised as white $1/f^0$ to red $1/f^2$ noise. In line with the snapshots of Figure \ref{fig:Snapshot}, `whiter' noises require greater levels of tolerance and result in accordingly lower fitness. Points show medians and error bars show 2.5\textsuperscript{th} and 97.5\textsuperscript{th} percentiles, each over the last 130,000 iterations of a given run.}
    \label{fig:Tolerance}
\end{figure}

\begin{figure}[h!]
    \includegraphics[width=\linewidth]{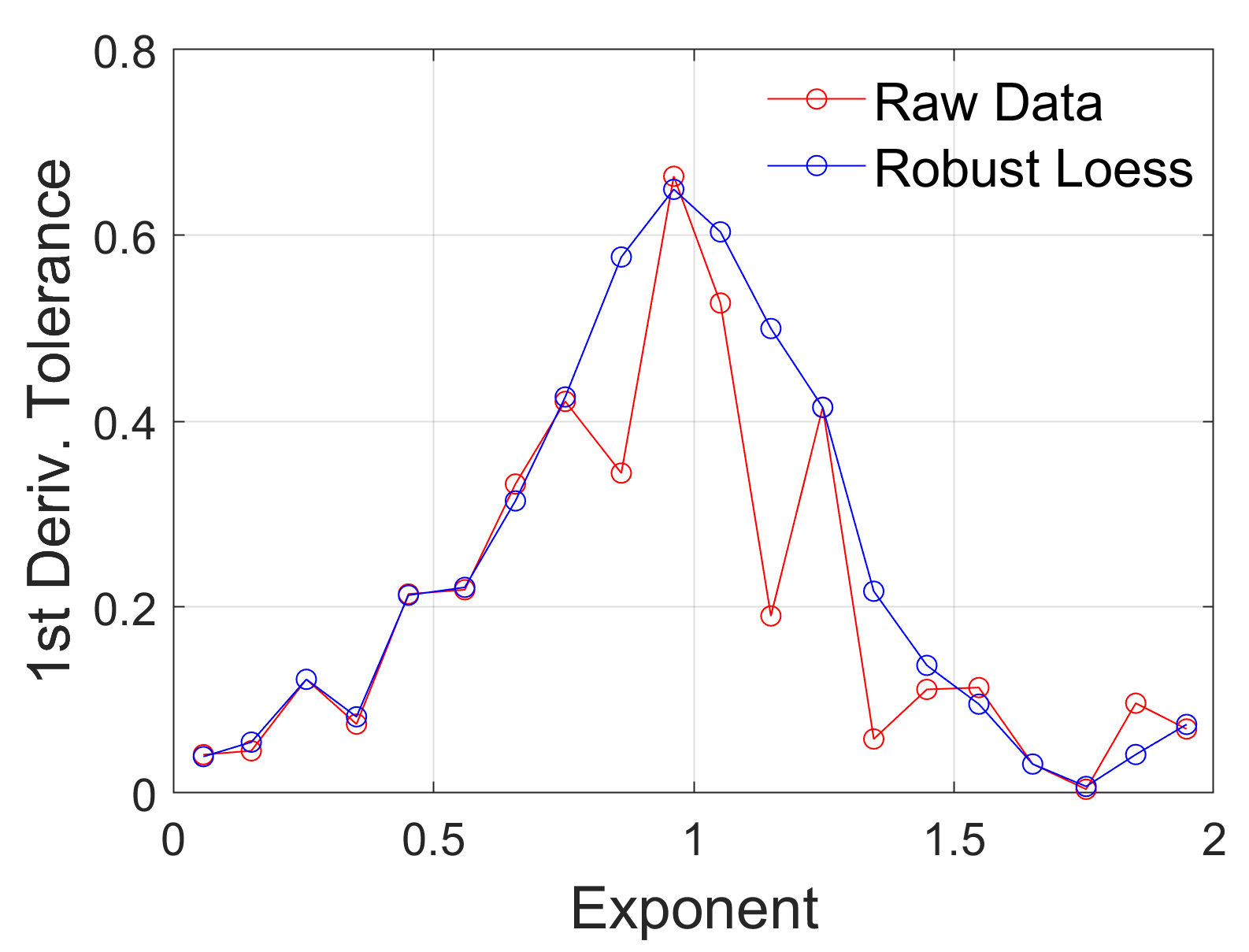}
    \caption{A robust LOESS curve fitted against the 1\textsuperscript{st} derivative of raw tolerance data (as seen in Figure \ref{fig:Tolerance}). The fitted curve demonstrating that the rate of change in tolerance against noise exponent peaks at pink $1/f$ noise.}
    \label{fig:LOESS}
\end{figure}
Coloured noise environments are generated via the Inverse Fast Fourier Transform (henceforth IFFT), which provides truer approximations to real coloured noises than more commonly used autoregressive processes. Generation via the IFFT involves the following steps:\begin{enumerate}
    \item A power spectral density (PSD) function is generated according to  $1/f^{\beta}$, with $f=(t/2s)/n$; here $t=1,2,...,m$, $s$ is an arbitrary sampling frequency (here $s=1$) and $n$ is equal to the length of the vector $t$. For greatest speed and accuracy when using the IFFT, $n$ should be a power of 2;
    \item The PSD is then converted from power to amplitude, using $ASD=sqrt(2*PSD)$;
    \item $ASD$ is then doubled in length by adding a version of the $ASD$ vector rotated by 180 degrees to the bottom of the existing vector;
    \item A vector of random phase angles $\theta$ on the interval $(0,2\pi)$ of length $2n$ is generated;
    \item The final vector $V=ASD*exp(i\theta)$, with $i=\sqrt{-1}$, is passed to the IFFT, with only the real component of the resulting complex vector retained. Note that the final coloured noise time series is of length $2n$.
\end{enumerate}
Here we use $n=65,536$, leading to coloured noise series of length 131,072. We generate coloured noises with $\beta$ values from 0 to 2 in increments of 0.1. Each time series is then scaled to have unit variance by dividing each value by the existing variance (as per \cite{Wichmann2005}). Overall median tolerances (median $\sigma$) are calculated by taking the 50\textsuperscript{th} percentiles of the distributions of all agents over the last 130,000 iterations of each run. Similarly, 95 confidence intervals are calculated as the 2.5\textsuperscript{th} and 97.5\textsuperscript{th} percentiles of these distributions. Where we plot median tolerances against $\beta$ values, we use the actual $\beta$ values calculated from log-spectral FFT analyses of the generated time series, as these sometimes differ very slightly from the desired $\beta$ values embodied in the original PSD.

\section{Results}
2,000-iteration snapshots of example runs are shown for $1/f^0$ (white noise), $1/f^1$ (pink noise), and $1/f^2$ (red noise) in Figure \ref{fig:Snapshot}. It is immediately clear from these examples that agents are more successful at tracking the simulated environments as $\beta$ increases (i.e. as the noise is `reddened'). Selection for greater tolerance occurs when agents are unable to closely track an environment, and as such there are greater tolerance values in `whiter' environments. Figure \ref{fig:Tolerance} (top) makes this pattern explicit; greater tolerance is required in `whiter' environments. As would be expected, greater tolerance implies lower fitness (Figure \ref{fig:Tolerance}, bottom), as agents are forced to generalise in response to a widely fluctuating environment rather than specialise on a relatively stable one. There is thus a strong stimulus towards greater tolerance, generalism, or flexibility in `whiter' environments, whilst agents in `redder' environments can be \emph{relatively} specialised.

Evolution under $1/f^1$ (pink) noise provides us with an interesting intermediary between white and red noise that, given the assertion of \cite{Halley1996} that pink noise can be viewed as an appropriate model of typical environmental noise, requires some exploration. A local polynomial regression method, known as robust locally estimated scatterplot smoothing (LOESS), is applied to the 1\textsuperscript{st} derivative of the raw tolerance data (see Figure \ref{fig:LOESS}, raw data summarised in Figure \ref{fig:Tolerance}), in order to smooth the raw results to approximate the relationship between the environmental exponent (noise colour) and the rate of change in the evolved tolerance. The robust LOESS curve demonstrates that, as the environment moves from white $1/f^0$ to pink $1/f$ noise, the rate of change in evolved tolerance increases to its peak, falling away as the environment reddens beyond pink noise. This demonstrates that pink $1/f$ noise is not just the central point in terms of exponent, but also the pivot point between two relatively stable evolutionary states: high tolerance in white noised environments, and and low tolerance in red noised environments. 

\section{Discussion}

The above results, albeit achieved with a simple model, are sufficient to demonstrate that the colour of noise chosen to represent the environment to which agents are adapting has a profound influence on model output. Two of the most frequently used models for environmental noise, simple Gaussian (white noise) and Brownian (red noise) signals, are at opposite ends of a wide spectrum of variation and lead to widely divergent results.

As well as serving as a warning of the dependence of results on the model used to simulate environmental fluctuations, the results shown in Figure \ref{fig:Tolerance} reveal some more specific features of the evolutionary response to fluctuations of different `colours'. The fact that `whiter' environments require greater levels of tolerance is in line with much recent theorising and research in numerous disciplines (e.g. \cite{Godfrey-Smith1996,Ash&Gallup2007,Sol2009}). In particular, the last two decades have seen a shift in explanations of the palaeoenvironmental impacts on human evolution from a focus on `habitat-specific' or `homogeneity' hypotheses to a focus on `heterogeneity' hypotheses. The preeminent example of the former is the `savanna hypothesis' (e.g. \cite{Dart1925,Dominguez-Rodrigo2014}); this posits that a long-term cooling and drying trend over the course of the Miocene led to a decrease in forest cover, a concomitant increase in savanna, and in turn to many of the adaptations that are seen as central to human evolution. The evolution of bipedalism, the emergence of tool use, and the increasing reliance on large game hunting have all been explained via this hypothesis.

Heterogeneity hypotheses, by contrast, focus on the constantly shifting environments that our ancestors (and other animals) would have been subject to over the course of their evolution. The exemplar is Potts' `variability selection hypothesis' \citep{Potts1996,Potts1998,Potts2013}, which stresses that organisms forced to adapt to widely varying conditions will of necessity develop greater tolerance or versatility, and will therefore be more able to deal with novel and unpredictable environments that they encounter in the future. Potts' ideas are consistent with the widely established `geometric mean effect' in evolutionary biology, by which those organisms (or genes) with the lowest variance in fitness over extended periods rather than those with the highest instantaneous fitness at a given time will be more successful (e.g. \cite{Lewontin&Cohen1969,Philippi&Seger1989,Simons2002}).

The contrast between long-term environmental change and short-term environmental variability is of course to a certain extent scale-dependent. Attempts to empirically decompose palaeoenvironmental time series into these two components have generally involved de-trending the time series via a polynomial of (arbitrary) order $k$; the polynomial is then regarded as the change and the residuals as the variability. This approach is subjective and unsatisfactory, but does point to a general - and valuable - understanding that the variability component varies symmetrically around the change component, that it is not the product of recognised periodic fluctuations (e.g. Milankovitch cycles), and that it might therefore resemble something like white noise.

If we view variability as being essentially the white noise component of a signal, then our results (Figure \ref{fig:Tolerance}) are fully consistent with the work of Potts and others. As the amplitude of variability increases, so the need for greater tolerance or versatility increases. The cumulative effects of this process, in response to the steadily increasing variability evident in empirical palaeoenvironmental signals over the past five million years (e.g. \cite{Lisiecki&Raymo2005}), could form the basis for the steadily increasing flexibility that we observe in hominin behaviour over that period.

Natural selection occurs primarily at a generational timescale; this suggests both further expansions of the above model and potential explanations for the extreme versatility of hominin species. The results presented above rely on two parameters that are set arbitrarily here: the mutation rate and the proportion of agents replaced each iteration. Since half the population is replaced each iteration the \emph{average} generation time of an agent in the simulation is 2 iterations. When we increase the mutation rate or the proportion replaced each iteration (the latter would decrease average generation time), populations are better able to track even white noise environments, and thus evolve lower levels of tolerance or versatility. This finding suggests that organisms with shorter generation times (or higher mutation rates) are more able to track changing environments, and should therefore exhibit fewer signs of versatility. This may indeed be the case (see \cite{Grove2017}), and it is certainly true that animals with shorter generation times have, on average, smaller brains when compared to longer-lived sister taxa \citep{Grove2017}. Since the brain is the principle organ of behavioural versatility, it is likely that there is a relationship between life-history and behavioural versatility \citep{Grove2020}. There are a range of confounding relationships and collinearities - not least the fact the recombination events, a major generator of new genotypic variation, occur less frequently in animals with longer generation times - but this appears to be a profitable line of enquiry for both evolutionary ecology and artificial life.

The above results also begin to shed some light on why pink noise might be special where evolutionary dynamics is concerned. It has already been noted by \cite{Inchausti2002} that animal population dynamics typically exhibit $1/f$ `pink' noise at a level that appears to be greater than would be expected from the environmental noise that the population was exposed to, with \cite{Halley1996} proposing that pink noise can be viewed as an appropriate model of typical environmental noise. Both the tolerance and fitness plots in Figure \ref{fig:Tolerance} exhibit a sigmoidal relationship with noise colour, with Figure \ref{fig:LOESS} going on to show that the rate of change of the 1\textsuperscript{st} derivative of the raw tolerance data achieves its maximum rate of change when the noise exponent $=1$. Pink noise being the pivot around which a change from high tolerance (under white noise) to low tolerance (under red noise) occurs suggests that populations which evolve under pink noise, or produce emergent evolutionary dynamics that exhibit pink noise, might be better placed to move in to new environmental niches or out-compete less or more tolerant populations if the external environment exhibits unexpected levels of variability. Pink noise, with its proportional input from all frequencies, its long memory, and its non-stationarity \citep{Halley2004b}, seems to offer a balance between the need for tolerance of large potential changes and the need to track the environmental mean (see Figure \ref{fig:Snapshot}). This balance between the need for tolerance and the need to achieve high relative fitness in the environmental conditions encountered most frequently is clearly demonstrated in Figure \ref{fig:LOESS}.

\section{Conclusions}

Artificial Life research often includes noise, either built in to the evolutionary dynamics, or as part of some external environment. However, despite attempts to explore naturally inspired evolutionary phenomena, the noise used (especially when creating external environments) is rarely inspired by the kinds of time series actually observed in nature. Here, using a simple evolutionary algorithm, we have demonstrated that the resulting tolerance of an evolving population is heavily dependant on the `colour' of the environmental time series, with `whiter' time series inducing greater tolerance, but at the cost of poor tracking of the environmental mean, and `redder' noise inducing higher levels of specialisation. Pink noise, with its proportional contribution from all frequencies, its long memory, and its non-stationarity, has been touted as an appropriate model of typical environmental noise \citep{Halley1996, Halley2004b}. We observe that pink noise induces a balance between generalism and specialism, providing the transition point between these two strategies. The results presented here also feed in to a wider discussion of the palaeoenvironmental impacts on human evolution, with our results further demonstrating that the `whitening' of the environmental time series requires the emergence of heterogeneous rather than homogeneous forms of adaptation. Ultimately we believe coloured $1/f^\beta$ noise is a more appropriate model for environmental variability than those models currently employed within Artificial Life.

\section{Acknowledgements}
This work is supported by Keele University, Faculty of Natural Sciences, Research Development Funding.
\bibliographystyle{apalike}
\bibliography{colourednoise.bbl}
\end{document}